\documentstyle[12pt,prd,aps,epsfig,preprint]{revtex}
\begin{document} \draft
\date{\today}
\tightenlines
\title{An analysis of $f_0-\sigma$ mixing in light cone QCD sum rules }

\author{A. Gokalp~\thanks{agokalp@metu.edu.tr}, Y. Sarac and
        O. Yilmaz~\thanks{oyilmaz@metu.edu.tr}}
\address{ {\it Physics Department, Middle East Technical University,
06531 Ankara, Turkey}}

\maketitle

\begin{abstract}
We investigate $f_0-\sigma$ mixing in the framework of light cone
QCD sum rules and by employing the experimental results about the
decay widths and the masses of these scalar mesons we estimate the
scalar mixing angle by using $\sigma$ meson data and $f_0$ meson
data. The two values we thus obtain of the scalar mixing angle are
not entirely consistent with each other, possibly indicating that
the structure of these mesons cannot be simple quark-antiquark
states.
\end{abstract}

\thispagestyle{empty} ~~~~\\

\newpage
\setcounter{page}{1}

The light scalar mesons have been the subject of continuous
interest in hadron spectroscopy. In addition to well established
isoscalar $f_0(980)$ and isovector $a_0(980)$ scalar mesons, the
recent experiments and theoretical analysis find evidence for the
existence of an isoscalar $\sigma(600)$ \cite{R1}, and an
isodoublet $\kappa(900)$ scalar mesons \cite{R2}. Although the
structure of these light scalar mesons have not been unambiguously
determined yet \cite{R3}, the possibility may be suggested that
these nine scalar mesons below 1 GeV may form a scalar SU(3)
flavor nonet \cite{R4}. In this case, one would expect a mixing in
the scalar sector similar to that occurs in the pseudoscalar
sector. One thus expresses the physical $\sigma(600)$ and
$f_0(980)$ states as linear combinations of isoscalar
$s\overline{s}$ and $(u\overline{u}+d\overline{d})/\sqrt{2}$
states as
\begin{eqnarray}\label{e1}
  \sigma &=&
  \cos\theta_s~\frac{u\overline{u}+d\overline{d}}{\sqrt{2}}-\sin\theta_s~s\overline{s}
  \nonumber \\
  f_0 &=&
  \sin\theta_s~\frac{u\overline{u}+d\overline{d}}{\sqrt{2}}+\cos\theta_s~s\overline{s}~~,
\end{eqnarray}
where $\theta_s$ is the scalar mixing angle.

In this work, we estimate the scalar mixing angle $\theta_s$ by
employing light cone QCD sum rules. The method has been used for
the calculation of hadronic coupling constants and for the studies
of hadronic properties \cite{R5}. We study the hadronic decays
$f_0\rightarrow \pi\pi$ and $\sigma\rightarrow \pi\pi$ within the
framework of this method and by utilizing the experimental data
about these decays and about the masses and the widths of $f_0$
and $\sigma$ mesons we deduce the scalar mixing angle $\theta_s$.

In order to study $S\pi\pi$-vertex, where S denotes the scalar
meson $f_0$ or $\sigma$, we consider the two point correlation
function
\begin{equation}\label{e2}
  T_{\mu}(p,q)=i\int d^{4}x ~e^{ip\cdot x}
  <\pi^+(q)|T\{j_\mu^{\pi^-}(x)j^S(0)\}|0>~~,
\end{equation}
where $p$ and $j_{\mu}^{\pi^-}$ are the four-momentum and the
interpolating current for $\pi^-$ meson, $j^S$ is the
interpolating current for scalar $f_0$ or $\sigma$ meson, and $q$
is the four-momentum of $\pi^+$ state. The interpolating currents
in terms of quark fields are the axial vector
$j_{\mu}^{\pi^-}(x)=\overline{u}(x)\gamma_\mu\gamma_5 d(x)$ for
pion, and the scalar currents
$j^{f_0}(x)=(\sin\theta_s/\sqrt{2})[\overline{u}(x)u(x)+\overline{d}(x)d(x)]+\cos\theta_s~\overline{s}(x)s(x)$
for $f_0$ meson and
$j^{\sigma}(x)=(\cos\theta_s/\sqrt{2})[\overline{u}(x)u(x)+\overline{d}(x)d(x)]-\sin\theta_s~\overline{s}(x)s(x)$
for $\sigma$ meson. The scalar current $j^S$ for $f_0$ and
$\sigma$ meson is assumed to have a non-vanishing matrix element
between the vacuum and a scalar meson state, $<S\mid j^S\mid
0>=\lambda_S$, where $\lambda_S$ is called the overlap amplitude.
The overlap amplitude $\lambda_\sigma$ for $\sigma$ meson
\cite{R6} and the overlap amplitude $\lambda_{f_0}$ for $f_0$
meson \cite{R7} were previously determined by using two-point QCD
sum rules method.

The correlation function $T_{\mu}(p,q)$ can be written in terms of
two independent invariant functions T$_1$ and T$_2$ by using
Lorentz invariance as
\begin{equation}\label{e3}
  T_{\mu}(p,q)=iT_1\left((p+q)^2,~p^2\right)p_\mu+T_2\left((p+q)^2,~p^2\right)q_\mu~~.
\end{equation}
We consider the invariant function T$_1((p+q)^2,~p^2)$. In order
to construct the theoretical part of the sum rule in terms of QCD
degrees of freedom  we calculate the function T$_1$  by evaluating
the correlation function $T_\mu$ in the deep Euclidean region,
where $(p+q)^2$ and $p^2$ are large and negative, as an expansion
near the light cone $x^2=0$. This expansion involves matrix
elements of non-local quark-gluon operators between pion and
vacuum states which defines pion distribution amplitudes of
increasing twist. We retain terms up to twist four accuracy since
higher twist amplitudes are suppressed by powers of $1/[-(p+q)^2]$
or $1/[-(p)^2]$ and thus are known to make a small contribution
\cite{R8}. In our calculation we use the full light quark
propagator with both perturbative and nonperturbative
contributions which is given as \cite{R9}
\begin{eqnarray}\label{e4}
  iS(x,0)&=&<0|T\{\overline{q}(x)q(0)\}|0>\nonumber \\
         &=&i\frac{\not x}{2\pi^2x^4}-\frac{<\overline{q}q>}{12}-
         \frac{x^2}{192}m_0^2<\overline{q}q>\nonumber \\
         &~&-ig_s\frac{1}{16\pi^2}\int_0^1du\left\{
         \frac{\not{x}}{x^2}\sigma_{\mu\nu}G^{\mu\nu}(ux)
         -4iu\frac{x_\mu}{x^2}G^{\mu\nu}(ux)\gamma_\nu\right\}+...~~.
\end{eqnarray}
We note that the two-point correlation function $T_\mu$, and
therefore the invariant function $T_1$, have contributions coming
only from the first terms of the interpolating currents for $f_0$
or $\sigma$ mesons. In other words, since the pion does not have
any strangeness content only the
$(\sin\theta_s/\sqrt{2})(\overline{u}u+\overline{d}d)$ part of the
interpolating current for $f_0$ meson and the
$(\cos\theta_s/\sqrt{2})(\overline{u}u+\overline{d}d)$ part of the
interpolating current for $\sigma$ meson make any contribution in
the evaluation of time-ordered products of pion and scalar meson
currents in Eq. 2. We, therefore, in our calculation of the
theoretical part of the sum rule consider the scalar current as
$j^S(x)=[\overline{u}(x)u(x)+\overline{d}(x)d(x)]/\sqrt{2}$, and
denote the resulting expression for the invariant function $T_1$
as $T'_1$. After a straightforward computation and after
performing the Fourier transforms the invariant function $T'_1$ is
obtained to twist four accuracy as
\begin{eqnarray}\label{e4}
&&T'_1\left((p+q)^2,~p^2\right)=\frac{f_\pi M_\pi^2}{2m_q}
  \int_0^1 du \left\{\left[-\frac{1}{(p+q-uq)^2}~\varphi_{P\pi}(u)-
  \frac{1}{3}~\frac{(p+q-uq)\cdot q}{(p+q-uq)^4}~\varphi_{\sigma\pi}(u)\right]\right. \nonumber \\
  &&~~~~~~~~~~~~~~~~~~~~~~~~~~~~~+\left.\left[-\frac{1}{(p+uq)^2}~\varphi_{P\pi}(u)-
  \frac{1}{3}~\frac{(p+uq)\cdot
  q}{(p+uq)^4}~\varphi_{\sigma\pi}(u)\right]\right\}\nonumber \\
&&+f_{3\pi}\int_0^1 dv \left\{\int D\alpha_i\left[
  \frac{M_\pi^2}{[p+q(1-\alpha_1-v\alpha_3)]^4}~\varphi_{3\pi}(\alpha_i)
  -\frac{M_\pi^2}{[p+q(\alpha_1+v\alpha_3)]^4}~\varphi_{3\pi}(\alpha_i)\right]\right\}
\nonumber \\
&&+f_{3\pi}\int_0^12v~ dv \left\{\int D\alpha_i\left[
  -\frac{M_\pi^2}{[p+q(1-\alpha_1-v\alpha_3)]^4}~\varphi_{3\pi}(\alpha_i)
  +\frac{M_\pi^2}{[p+q(\alpha_1+v\alpha_3)]^4}~\varphi_{3\pi}(\alpha_i)\right]\right\}~.
\end{eqnarray}
In the above expression the functions $\varphi_{\sigma \pi}(u)$
and $\varphi_{p\pi}(u)$ are the twist 3 quark-antiquark pion
distribution amplitudes which are defined by the matrix elements
\cite{R10}
\begin{eqnarray}\label{e6}
<\pi^+(q)|\overline{u}(x)i\gamma_5 d(0)|0> =f_\pi \mu_\pi \int_0^1
du e^{iuq\cdot x} \varphi_{p\pi}(u)~~,
\end{eqnarray}
\begin{eqnarray}\label{e7}
<\pi^+(q)|\overline{u}(x)\sigma_{\mu\nu}\gamma_5 d(0)|0> =
 i\frac{f_\pi\mu_\pi}{6}\left(1-\frac{M_\pi^2}{\mu_\pi^2}\right)(q_\mu
 x_\nu-x_\mu q_\nu) \int_0^1 du e^{iuq\cdot x} \varphi_{\sigma\pi}(u)~~,
\end{eqnarray}
where $\mu_\pi=M_\pi^2/2m_q$ with $m_q=(m_u+m_d)/2$ is the twist 3
distribution amplitude normalization factor. Although in the
evaluation of the light quark propagator given in Eq. 4 we put
$m_u=m_d=0$, we like to note that $m_q$ in the normalization
factor $\mu_\pi$ is obtained by making use of the
Gell-Mann-Oakes-Renner relation \cite{R11} as
$m_q=-f_\pi^2M_\pi^2/4\langle \overline{q}q\rangle$. We work in
the fixed-point gauge $x^\mu A_\mu=0$, consequently the
path-ordered gauge factors $P~exp\left\{ig_s\int_0^1d\alpha ~x_\mu
A^\mu(\alpha x)\right\}$ which appear in between the quark fields
and which assure gauge invariance are not included in the matrix
elements \cite{R9}. The twist 3 quark-antiquark-gluon pion
distribution amplitude $\varphi_{3\pi}(\alpha_i)$ is defined as
\cite{R10}
\begin{eqnarray}\label{e8}
 <\pi^+(q)|\overline{u}(x)g_s\gamma_5\sigma_{\alpha\beta}G{\mu\nu}(vx)d(0)|0> &=&
 if_{3\pi}\left[(q_\mu q_\alpha g_{\nu\beta}-q_\nu
q_\alpha g_{\mu\beta})-(q_\mu q_\beta g_{\nu\alpha}-q_\nu q_\beta
g_{\mu\alpha})\right] \nonumber \\
&\times&\int D\alpha_i \varphi_{3 \pi}(\alpha_i) e^{iq\cdot
x(\alpha_1+v\alpha_3)}
\end{eqnarray}
where $D\alpha_i=d\alpha_1 d\alpha_2 d\alpha_3
\delta(1-\alpha_1-\alpha_2-\alpha_3)$. After performing the Borel
transformation with respect to  the variables $Q_1^2=-(p+q)^2$ and
$Q_2^2=-p^2$, we obtain the theoretical expression for the
invariant function $T'_1$ in the form
\begin{eqnarray}\label{e9}
 T'_1(M_1^2,~M_2^2)&=&\frac{f_\pi M_\pi^2 M^2}{2m_q} \left[\varphi_{p\pi}(u_{01})
  +\frac{1}{6}~\frac{d\varphi_{\sigma \pi}}{du}\mid_{u=u_{01}}\right]
\nonumber \\
  &+&f_{3\pi}M_\pi^2\int_0^{u_{01}}d\alpha_1\int_{u_{01}-\alpha_1}^{1-\alpha_1}\frac{d\alpha_3}{\alpha_3}~
\varphi_{3\pi}(\alpha_1,1-\alpha_1-\alpha_3,\alpha_3)\left(2\frac{u_{01}-\alpha_1}{\alpha_3}-1\right)
\nonumber \\
&+&\frac{f_\pi M_\pi^2 M^2}{2m_q} \left[\varphi_{p\pi}(u_{02})
  -\frac{1}{6}~\frac{d\varphi_{\sigma \pi}}{du}\mid_{u=u_{02}}\right]
\nonumber \\
  &-&f_{3\pi}M_\pi^2\int_0^{u_{02}}d\alpha_1\int_{u_{02}-\alpha_1}^{1-\alpha_1}\frac{d\alpha_3}{\alpha_3}~
\varphi_{3\pi}(\alpha_1,1-\alpha_1-\alpha_3,\alpha_3)\left(2\frac{u_{02}-\alpha_1}{\alpha_3}-1\right)
\end{eqnarray}
where M$_1^2$ and M$_2^2$ are the Borel parameters and
\begin{eqnarray}
  u_{01}=\frac{M_2^2}{M_1^2+M_2^2}~~~~,~~
u_{02}=\frac{M_1^2}{M_1^2+M_2^2}~~~~,~~
M^2=\frac{M_1^2M_2^2}{M_1^2+M_2^2} ~~. \nonumber
\end{eqnarray}

The correlation function  $T_\mu(p,q)$ satisfies a dispersion
relation, therefore we can represent the invariant function $T_1$
as
\begin{equation}\label{e10}
  T_1\left((p+q)^2,~p^2\right)=\int\int ds ds'~
  \frac{\rho^{had}(s,s')}{[s-(p+q)^2)](s'-p^2)}~~,
\end{equation}
where the hadronic spectral density $\rho^{had}(s,s')$ receives
contributions from the single-particle states and also from higher
resonances and continuum states. We therefore saturate this
dispersion relation by inserting a complete set of one
hadron-states into the correlation function and we consider the
single-particle $\pi$ and scalar meson S, where S denotes $f_0$ or
$\sigma$, states. This way we obtain the expression for the
invariant function $T_1$ in the form
\begin{eqnarray}\label{e11}
 T_1\left((p+q)^2,~p^2\right)=&&\frac{<0\mid j_{\mu}^{\pi^-}\mid \pi^+(p)>
   <\pi^+\pi^-\mid S> <S(p+q)|j^S\mid 0>}
  {\left[(p+q)^2-M_{S}^2\right]\left(p^2-M_\pi^2\right)} \nonumber \\
  &&+\int_{s_{0}} ds\int_{s'_{0}}ds'~\frac{\rho^{cont}(s,s')}{[s-(p+q)^2](s'-p^2)}~~,
\end{eqnarray}
where now the hadronic spectral density $\rho^{had}(s,s')$
includes the contributions of higher resonances and the hadronic
continuum which contribute in a domain ${\cal D}$ of the $(s,s')$
plane starting from two thresholds $s_0$ and $s'_0$. The matrix
element $<\pi^+\pi^-\mid S>$ defines the coupling constant
g$_{S\pi\pi}$
\begin{eqnarray}\label{e12}
<\pi^+(q)\pi^-(p)\mid S(p+q)>=g_{S\pi\pi}
\end{eqnarray}
and the current-particle matrix elements are given as
\begin{eqnarray}\label{e13}
<S(p+q)\mid j^S\mid 0>=\lambda_S~~,
\end{eqnarray}
\begin{eqnarray}\label{e14}
<0\mid j_{\mu}^{\pi^-}\mid \pi^+(p)>=if_\pi p_\mu~~.
\end{eqnarray}
The contribution coming from the continuum can be identified by
using global quark-hadron duality \cite{R12} with the QCD
contribution above the thresholds $s_0$ and $s'_0$. This way the
pole contribution in which the coupling constant g$_{S\pi\pi}$
appears is isolated. After performing the same Borel
transformation that was applied to the theoretical expression for
the invariant function $T'_1$ we obtain for the hadronic
representation of the same invariant function the result
\begin{eqnarray}\label{e15}
 T_1(M_1^2,M_2^2)&=&\lambda_S~f_\pi~ g_{S\pi\pi}~e^{-M^2_S/M_1^2}~e^{-M^2_\pi/M_2^2}
\nonumber \\
&+&\int_{s_{0}}ds\int_{s'_{0}}ds'~\rho^{cont}(s,s')~e^{-s/M_1^2}~e^{-s'/M_2^2}~~.
\end{eqnarray}
The sum rule for the coupling constant g$_{S\pi\pi}$ then follows
in accordance with the QCD sum rules method strategy by equating
the expressions $T_1(M_1^2,M_2^2)$ obtained for the invariant
function $T_1((p+q)^2,p^2)$ by theoretical calculation given in
Eq. 9 and by physical considerations which is given in Eq. 15. For
this purpose we note that the expression for $T'_1$ given in Eq. 9
should be multiplied by $\cos\theta_s$ to obtain the sum rule for
g$_{\sigma\pi\pi}$ and by $\sin\theta_s$ for the corresponding sum
rule for g$_{f_0\pi\pi}$. In order to do this we have to identify
the second term in Eq. 15 representing the contribution of the
continuum with a part of the term calculated theoretically in QCD.
We follow the prescription, for this so called the subtraction of
the continuum, for the cases where the Borel parameters
corresponding to channels with different mass scales cannot be
constrained to be equal \cite{R13,R14}. This prescription is based
on the observation that the distribution amplitudes
$\varphi_{p\pi}(u)$ and $\varphi_{\sigma \pi}(u)$ are polynomials
in $(1-u)$, therefore in order to compute their contribution in
the duality region ${\cal D}$ we can write
\begin{eqnarray}\label{e16}
 \varphi_{p\pi}(u)+\frac{1}{6}~\frac{d\varphi_{\sigma\pi}}{du}=\sum_{k=0}^{N}b_k(1-u)^k~~,
~~~~~~\varphi_{p\pi}(u)-\frac{1}{6}~\frac{d\varphi_{\sigma\pi}}{du}=\sum_{k=0}^{N}b'_k(1-u)^k~~.
\end{eqnarray}
Since the contribution of the twist 3 quark-antiquark-gluon term
in Eq. 9 is small, we thus affect the continuum subtraction in the
leading twist 3 quark-antiquark term. Therefore, we finally obtain
the sum rules for the coupling constants g$_{\sigma\pi\pi}$ and
g$_{f_0\pi\pi}$ in the form
\begin{eqnarray}\label{e17}
g_{\sigma\pi\pi}=\cos\theta_s~ g'_{\sigma\pi\pi}~~,~~~~~~
g_{f_0\pi\pi}=\sin\theta_s~ g'_{f_0\pi\pi}
\end{eqnarray}
where
\begin{eqnarray}\label{e18}
 g'_{S\pi\pi}&=&\frac{1}{\lambda_S}e^{M^2_S/M_1^2}e^{M^2_\pi/M_2^2}\left\{
 \frac{M^2M_\pi^2}{2m_q}\sum_{k=0}^{N}b_k\left(\frac{M^2}{M_2^2}\right)^k
\left[1-e^{-A}\sum_{i=0}^{k}\frac{A^i}{i!}+e^{-A}\frac{M^2M_\pi^2}{M_1^2M_2^2}\frac{A^{(k+1)}}{(k+1)!}\right]
\right. \nonumber \\
&&+\left.
\frac{f_{3\pi}M_\pi^2}{f_\pi}\int_0^{u_{01}}d\alpha_1\int_{u_{01}-\alpha_1}^{1-\alpha_1}\frac{d\alpha_3}{\alpha_3}~
\varphi_{3\pi}(\alpha_1,1-\alpha_1-\alpha_3,\alpha_3)\left(2\frac{u_{01}-\alpha_1}{\alpha_3}-1\right)\right.
\nonumber \\
&&+
\frac{M^2M_\pi^2}{2m_q}\sum_{k=0}^{N}b'_k\left(\frac{M^2}{M_1^2}\right)^k
\left[1-e^{-A}\sum_{i=0}^{k}\frac{A^i}{i!}+e^{-A}\frac{M^2M_\pi^2}{M_1^2M_2^2}\frac{A^{(k+1)}}{(k+1)!}\right]
 \nonumber \\
&&-\left.
\frac{f_{3\pi}M_\pi^2}{f_\pi}\int_0^{u_{02}}d\alpha_1\int_{u_{02}-\alpha_1}^{1-\alpha_1}\frac{d\alpha_3}{\alpha_3}~
\varphi_{3\pi}(\alpha_1,1-\alpha_1-\alpha_3,\alpha_3)\left(2\frac{u_{02}-\alpha_1}{\alpha_3}-1\right)\right\}~~,
\end{eqnarray}
where $A=s_0/M^2$ with $s_0$ the smallest continuum threshold.

In the numerical evaluation of the sum rule we use the twist 3
pion distribution amplitudes given by \cite{R10}
\begin{eqnarray}\label{e19}
 \varphi_{p\pi}(u)=1&+&\left(30\frac{f_{3\pi}}{\mu_\pi f_\pi}-\frac{5}{2}\frac{M_\pi^2}{\mu_\pi^2}\right)C_2^{1/2}(2u-1)
\nonumber \\
 &+&\left[-3\frac{f_{3\pi}\omega_{3\pi}}{\mu_\pi
 f_\pi}-\frac{27}{20}\frac{M_\pi^2}{\mu_\pi^2}(1+6a_2^\pi)\right]C_4^{1/2}(2u-1)
\end{eqnarray}
\begin{eqnarray}\label{e20}
\varphi_{\sigma \pi}(u)=
 6u\overline{u}\left\{1+\left[5\frac{f_{3\pi}}{\mu_\pi f_\pi}(1-\frac{1}{10}\omega_{3\pi})
-\frac{7}{20}\frac{M_\pi^2}{\mu_\pi^2}(1+\frac{12}{7}a_2^\pi)\right]C_2^{3/2}(2u-1)\right\}
\end{eqnarray}
\begin{eqnarray}\label{e21}
 \varphi_{3\pi}(\alpha_i)=360\alpha_1\alpha_2\alpha_3^2\left[1+\frac{\omega_{3\pi}}{2}(7\alpha_3-3)\right]
\end{eqnarray}
where $C_m^k(2u-1)$ are the Gegenbauer polynomials. The overlap
amplitudes $\lambda_S$ were determined previously by employing
two-point QCD sum rules method as $\lambda_\sigma=(0.17\pm
0.02)~~GeV^2$ \cite{R6} and $\lambda_{f_0}=(0.18\pm 0.02)~~GeV^2$
\cite{R7}. We also adopt the values of the parameters in the
distribution amplitudes at the renormalization scale 1 GeV as
$f_{3\pi}(1GeV)=0.0035~~GeV^2$, $\omega_{3\pi}(1GeV)=-2.88$, and
$a_2^\pi=0$ \cite{R10}. Moreover, we also use  the values
$\langle\overline{q}q\rangle(1GeV)=-(0.240~~GeV)^3$ and
$f_\pi=0.132~~GeV$ \cite{R13}. The mass of the $\sigma$ meson is
taken as M$_\sigma=(483\pm 31)~~MeV$ \cite{R1} in which
statistical and systematic errors are added in quadrature.

We then study the dependencies of the sum rules for the coupling
constants g$'_{\sigma\pi\pi}$ and g$'_{f_0\pi\pi}$ on the Borel
parameters $M_1^2$ and $M_2^2$ by considering independent
variations of these parameters. As for the continuum threshold
parameter  $s_0$ we use the value $s_0=1.2~~GeV^2$ for $\sigma$
meson \cite{R6} and $s_0=1.1~~GeV^2$ for $f_0$ meson \cite{R7}
that were obtained in the calculation of the overlap amplitudes
$\lambda_\sigma$ and $\lambda_{f_0}$ using two-point QCD sum rules
method. We find that the sum rules are quite stable for the
variation of the Borel parameters in the ranges $2\leq M_1^2\leq
6~~GeV^2$ and $0.4\leq M_2^2\leq 0.8~~GeV^2$. The variation of the
coupling constant g$'_{\sigma\pi\pi}$ as a function of Borel
parameters $M_1^2$ and $M_2^2$ is shown in Fig. 1 and that of
g$'_{f_0\pi\pi}$ is shown in Fig. 2. The values of the coupling
constants thus can be obtained by choosing the value of $M_1^2$
about at the middle of the stability region as $3.2\leq
g'_{\sigma\pi\pi}\leq 3.9~~GeV$ and $3.4\leq g'_{f_0\pi\pi}\leq
3.8~~GeV$.

On the other hand, the decay widths of $\sigma$ and $f_0$ mesons
can be calculated in terms of the coupling constants
g$_{\sigma\pi\pi}$ and g$_{f_0\pi\pi}$ which are defined through
the matrix elements given in Eq. 12. The decay width $S\rightarrow
\pi^+\pi^-$ can thus be obtained in terms of the coupling constant
g$_{S\pi\pi}$ as
\begin{eqnarray}\label{e22}
  \Gamma(S\rightarrow \pi^+\pi^-)=\frac{1}{8\pi M_S^2}g^2_{S\pi\pi}~p^*~~,
\end{eqnarray}
where $p^*=\sqrt{M_S^2-(M_{\pi^+}+M_{\pi^-})^2}/2$ is the
magnitude of the three-momentum of the either of the pions in the
C.M. frame. The experimental width
\begin{eqnarray}\label{e23}
 \Gamma^{tot}(S\rightarrow\pi\pi)=\frac{3}{2}~\Gamma(S\rightarrow\pi^+\pi^-)=3~\Gamma(S\rightarrow\pi^0\pi^0)
\end{eqnarray}
is given as $\Gamma_{f_0}=40-100$ MeV with $M_{f_0}=(980\pm 10)$
MeV \cite{R15} for $f_0$ meson, and a best fit to the Dalitz plot
of $D^+\rightarrow\pi^+\sigma\rightarrow\pi^+\pi^+\pi^-$ decay in
E791 Collaboration experiment \cite{R1} results in
$M_{\sigma}=(483\pm 31)$ MeV and $\Gamma_\sigma=(338\pm 48)$ MeV
for $\sigma$ meson. If we use these experimental values we then
obtain the coupling constants g$_{S\pi\pi}$ as
g$_{\sigma\pi\pi}=(2.6\pm 0.2)~~GeV$ and g$_{f_0\pi\pi}=(1.6\pm
0.8)~~GeV$. The scalar mixing angle $\theta_s$ can then be
obtained by using Eq. 17 and utilizing on the one hand the values
of the coupling constants g$_{S\pi\pi}$ determined from
experimental results and on the other hand the values of
g$'_{S\pi\pi}$ estimated by light cone QCD sum rules method. This
way we determine the scalar mixing angle $\theta_s$ from $\sigma$
meson results as $\theta_s=(41\pm 11)^o$ and from $f_0$ meson
results as $\theta_s=(27\pm 13)^o$. These two values do not
exclude each other, since they have a small overlap region, but
given the large difference in the average values of the scalar
mixing angle $\theta_s$ determined using $\sigma$ meson and $f_0$
meson properties we may assert that our results indicate that the
structure of $\sigma$ and $f_0$ mesons cannot be simple
quark-antiquark states.

It should be mentioned that the previous determinations of the
scalar mixing angle $\theta_s$ through experimental data coming
from charmed meson decays also lead to a somewhat inconsistent
picture. The $J/\psi\rightarrow f_0(980)\phi$ and
$J/\psi\rightarrow f_0(980)\omega$ decays were used to estimate
the mixing angle \cite{R16} and the result $\theta_s=(34\pm 6)^o$
was obtained, which is consistent with our results. The analysis
of the experimental results $D_S^+\rightarrow f_0(980)\pi^+$ and
$D_S^+\rightarrow \phi\pi^+$ \cite{R17} on the other hand yields
the mixing angle in the range $35^o\leq \theta_s\leq 55^0$. In
these analysis, it is the mixing angle in $f_0$ meson state that
is determined. However, such a mixing angle implies that $\sigma$
meson also has a strange component, therefore the decay
$D_S^+\rightarrow \sigma\pi^+$ should also occur, but this decay
is not observed to make any contribution to the $D_S^+\rightarrow
\pi^+\pi^+\pi^-$ decay by the E791 Collaboration \cite{R18}.
Therefore, this apparent inconsistency about the scalar mixing
angle may be taken to indicate that the structure of scalar mesons
$\sigma$ and $f_0$ are more complicated than simple
quark-antiquark states, a conclusion that is also supported by our
calculation.


\newpage

\begin{figure}\hspace{-0.5cm}
\epsfig{figure=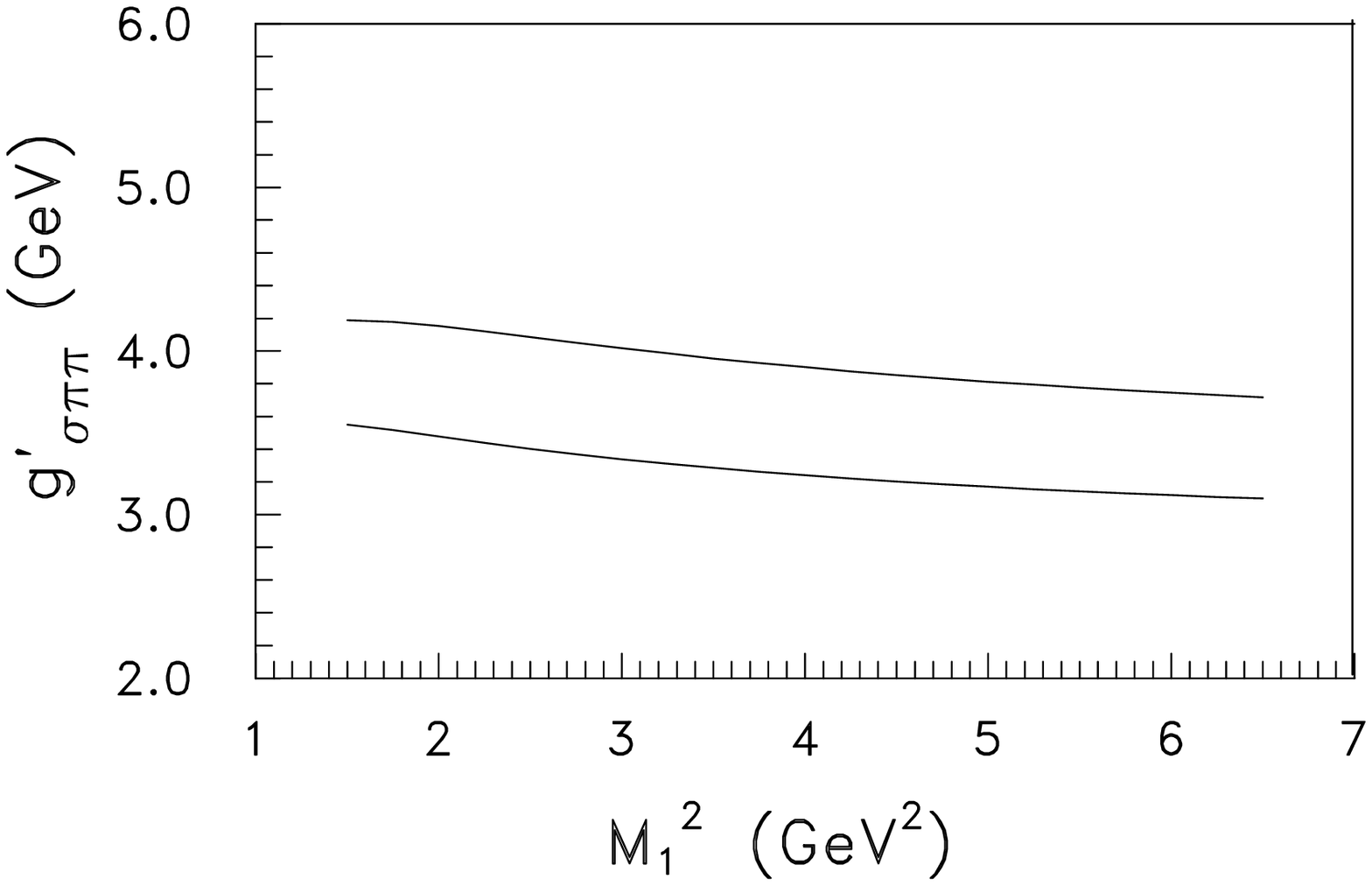,width=15cm,height=18cm} \vspace*{-3.5cm}
\caption{The coupling constant g$'_{\sigma\pi\pi}$ as a function
of the Borel parameter $M_1^2$ for different values of the Borel
parameter $M_2^2$. The curves denote the limits of the stability
region.} \label{fig1}
\end{figure}
\begin{figure}\hspace{-0.5cm}
\epsfig{figure=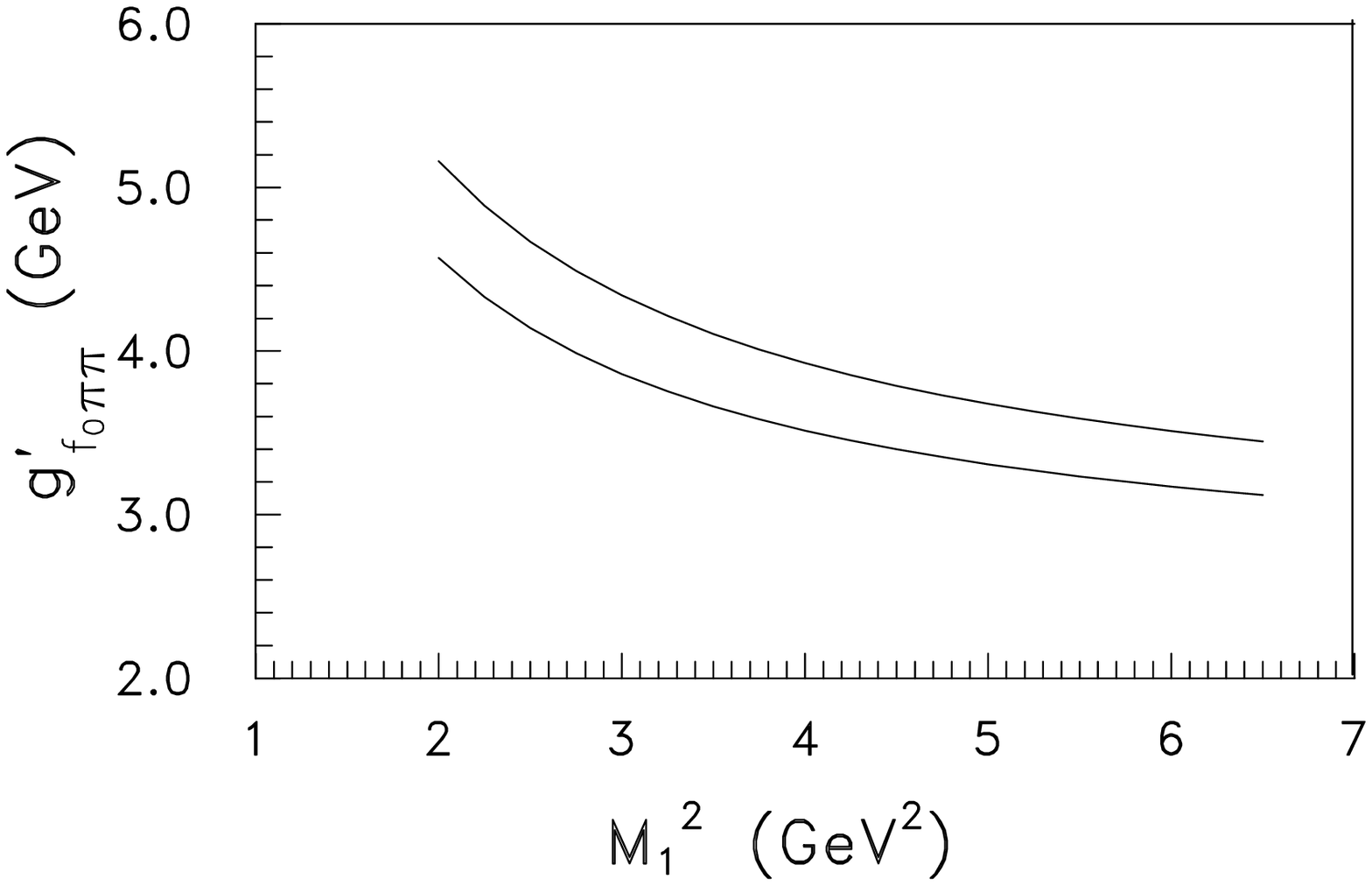,width=15cm,height=18cm} \vspace*{-3.5cm}
\caption{The coupling constant g$'_{f_0\pi\pi}$ as a function of
the Borel parameter $M_1^2$ for different values of the Borel
parameter $M_2^2$. The curves denote the limits of the stability
region.} \label{fig1}
\end{figure}

\end{document}